\documentclass[a4paper]{article}
\usepackage[english]{babel}
\usepackage{mathpazo}
\usepackage{amsmath,amssymb}
\usepackage{a4wide}
\usepackage{xspace}
\usepackage{graphicx}
\usepackage[margin=1.5ex,bf]{caption}
\usepackage{units}
\usepackage{booktabs}
\usepackage{xcolor}
\definecolor{darkred}{rgb}{0.3,0,0}
\definecolor{darkblue}{rgb}{0,0,0.3}
\definecolor{firebrick}{rgb}{0.5,0.125,0.125}
\definecolor{darkgreen}{rgb}{0,0.3,0}
\usepackage{hyperref}
\hypersetup{colorlinks=true,linkcolor=firebrick,citecolor=darkgreen,urlcolor=darkblue}
\usepackage{enumitem}
\usepackage{wrapfig}
\usepackage{longtable}
\usepackage{appendix}
\usepackage{multirow}
\usepackage{tabularx}

\begin{document}

\begin{center}
{\bf\large The Pierre Auger Observatory Open Data}
\end{center}

\begin{wrapfigure}[9]{l}{0.12\linewidth}
\vspace{-2.9ex}
\includegraphics[width=0.98\linewidth]{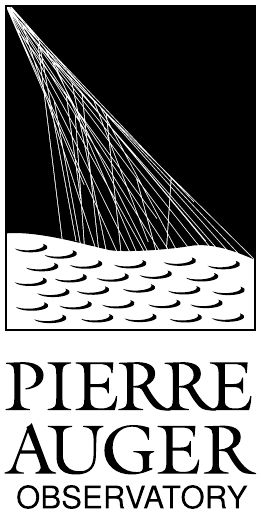}
\end{wrapfigure}
\begin{sloppypar}\noindent
A.~Abdul Halim$^{13}$,
P.~Abreu$^{70}$,
M.~Aglietta$^{53,51}$,
I.~Allekotte$^{1}$,
K.~Almeida Cheminant$^{78,77}$,
A.~Almela$^{7,12}$,
R.~Aloisio$^{44,45}$,
J.~Alvarez-Mu\~niz$^{76}$,
A.~Ambrosone$^{44}$,
J.~Ammerman Yebra$^{76}$
% ...

\end{sloppypar}
\begin{center}
\par\noindent
\textbf{The Pierre Auger Collaboration}
\end{center}

\vspace{1ex}
\begin{center}
\rule{0.1\columnwidth}{0.5pt}
\raisebox{-0.4ex}{\scriptsize$\bullet$}
\rule{0.1\columnwidth}{0.5pt}
\end{center}

\vspace{1ex}
\begin{description}[labelsep=0.2em,align=right,labelwidth=0.7em,labelindent=0em,leftmargin=2em,noitemsep,before={\renewcommand\makelabel[1]{##1 }}]
\item[$^{1}$] Centro At\'omico Bariloche and Instituto Balseiro (CNEA-UNCuyo-CONICET), San Carlos de Bariloche, Argentina
\item[$^{2}$] Departamento de F\'\i{}sica and Departamento de Ciencias de la Atm\'osfera y los Oc\'eanos, FCEyN, Universidad de Buenos Aires and CONICET, Buenos Aires, Argentina
\item[$^{3}$] IFLP, Universidad Nacional de La Plata and CONICET, La Plata, Argentina
\item[$^{4}$] Instituto de Astronom\'\i{}a y F\'\i{}sica del Espacio (IAFE, CONICET-UBA), Buenos Aires, Argentina
\item[$^{5}$] Instituto de F\'\i{}sica de Rosario (IFIR) -- CONICET/U.N.R.\ and Facultad de Ciencias Bioqu\'\i{}micas y Farmac\'euticas U.N.R., Rosario, Argentina
\item[$^{6}$] Instituto de Tecnolog\'\i{}as en Detecci\'on y Astropart\'\i{}culas (CNEA, CONICET, UNSAM), and Universidad Tecnol\'ogica Nacional -- Facultad Regional Mendoza (CONICET/CNEA), Mendoza, Argentina
\item[$^{7}$] Instituto de Tecnolog\'\i{}as en Detecci\'on y Astropart\'\i{}culas (CNEA, CONICET, UNSAM), Buenos Aires, Argentina
\item[$^{8}$] International Center of Advanced Studies and Instituto de Ciencias F\'\i{}sicas, ECyT-UNSAM and CONICET, Campus Miguelete -- San Mart\'\i{}n, Buenos Aires, Argentina
\item[$^{9}$] Laboratorio Atm\'osfera -- Departamento de Investigaciones en L\'aseres y sus Aplicaciones -- UNIDEF (CITEDEF-CONICET), Argentina
\item[$^{10}$] Observatorio Pierre Auger, Malarg\"ue, Argentina
\item[$^{11}$] Observatorio Pierre Auger and Comisi\'on Nacional de Energ\'\i{}a At\'omica, Malarg\"ue, Argentina
\item[$^{12}$] Universidad Tecnol\'ogica Nacional -- Facultad Regional Buenos Aires, Buenos Aires, Argentina
\item[$^{13}$] University of Adelaide, Adelaide, S.A., Australia
\item[$^{14}$] Universit\'e Libre de Bruxelles (ULB), Brussels, Belgium
\item[$^{15}$] Vrije Universiteit Brussels, Brussels, Belgium
\item[$^{16}$] Centro Brasileiro de Pesquisas Fisicas, Rio de Janeiro, RJ, Brazil
\item[$^{17}$] Centro Federal de Educa\c{c}\~ao Tecnol\'ogica Celso Suckow da Fonseca, Petropolis, Brazil
\item[$^{18}$] Instituto Federal de Educa\c{c}\~ao, Ci\^encia e Tecnologia do Rio de Janeiro (IFRJ), Brazil
\item[$^{19}$] Universidade de S\~ao Paulo, Escola de Engenharia de Lorena, Lorena, SP, Brazil
\item[$^{20}$] Universidade de S\~ao Paulo, Instituto de F\'\i{}sica de S\~ao Carlos, S\~ao Carlos, SP, Brazil
\item[$^{21}$] Universidade de S\~ao Paulo, Instituto de F\'\i{}sica, S\~ao Paulo, SP, Brazil
\item[$^{22}$] Universidade Estadual de Campinas (UNICAMP), IFGW, Campinas, SP, Brazil
\item[$^{23}$] Universidade Estadual de Feira de Santana, Feira de Santana, Brazil
\item[$^{24}$] Universidade Federal de Campina Grande, Centro de Ciencias e Tecnologia, Campina Grande, Brazil
\item[$^{25}$] Universidade Federal do ABC, Santo Andr\'e, SP, Brazil
\item[$^{26}$] Universidade Federal do Paran\'a, Setor Palotina, Palotina, Brazil
\item[$^{27}$] Universidade Federal do Rio de Janeiro, Instituto de F\'\i{}sica, Rio de Janeiro, RJ, Brazil
\item[$^{28}$] Universidad de Medell\'\i{}n, Medell\'\i{}n, Colombia
\item[$^{29}$] Universidad Industrial de Santander, Bucaramanga, Colombia
\item[$^{30}$] Charles University, Faculty of Mathematics and Physics, Institute of Particle and Nuclear Physics, Prague, Czech Republic
\item[$^{31}$] Institute of Physics of the Czech Academy of Sciences, Prague, Czech Republic
\item[$^{32}$] Palacky University, Olomouc, Czech Republic
\item[$^{33}$] CNRS/IN2P3, IJCLab, Universit\'e Paris-Saclay, Orsay, France
\item[$^{34}$] Laboratoire de Physique Nucl\'eaire et de Hautes Energies (LPNHE), Sorbonne Universit\'e, Universit\'e de Paris, CNRS-IN2P3, Paris, France
\item[$^{35}$] Univ.\ Grenoble Alpes, CNRS, Grenoble Institute of Engineering Univ.\ Grenoble Alpes, LPSC-IN2P3, 38000 Grenoble, France
\item[$^{36}$] Universit\'e Paris-Saclay, CNRS/IN2P3, IJCLab, Orsay, France
\item[$^{37}$] Bergische Universit\"at Wuppertal, Department of Physics, Wuppertal, Germany
\item[$^{38}$] Karlsruhe Institute of Technology (KIT), Institute for Experimental Particle Physics, Karlsruhe, Germany
\item[$^{39}$] Karlsruhe Institute of Technology (KIT), Institut f\"ur Prozessdatenverarbeitung und Elektronik, Karlsruhe, Germany
\item[$^{40}$] Karlsruhe Institute of Technology (KIT), Institute for Astroparticle Physics, Karlsruhe, Germany
\item[$^{41}$] RWTH Aachen University, III.\ Physikalisches Institut A, Aachen, Germany
\item[$^{42}$] Universit\"at Hamburg, II.\ Institut f\"ur Theoretische Physik, Hamburg, Germany
\item[$^{43}$] Universit\"at Siegen, Department Physik -- Experimentelle Teilchenphysik, Siegen, Germany
\item[$^{44}$] Gran Sasso Science Institute, L'Aquila, Italy
\item[$^{45}$] INFN Laboratori Nazionali del Gran Sasso, Assergi (L'Aquila), Italy
\item[$^{46}$] INFN, Sezione di Catania, Catania, Italy
\item[$^{47}$] INFN, Sezione di Lecce, Lecce, Italy
\item[$^{48}$] INFN, Sezione di Milano, Milano, Italy
\item[$^{49}$] INFN, Sezione di Napoli, Napoli, Italy
\item[$^{50}$] INFN, Sezione di Roma ``Tor Vergata'', Roma, Italy
\item[$^{51}$] INFN, Sezione di Torino, Torino, Italy
\item[$^{52}$] Istituto di Astrofisica Spaziale e Fisica Cosmica di Palermo (INAF), Palermo, Italy
\item[$^{53}$] Osservatorio Astrofisico di Torino (INAF), Torino, Italy
\item[$^{54}$] Politecnico di Milano, Dipartimento di Scienze e Tecnologie Aerospaziali , Milano, Italy
\item[$^{55}$] Universit\`a del Salento, Dipartimento di Matematica e Fisica ``E.\ De Giorgi'', Lecce, Italy
\item[$^{56}$] Universit\`a dell'Aquila, Dipartimento di Scienze Fisiche e Chimiche, L'Aquila, Italy
\item[$^{57}$] Universit\`a di Catania, Dipartimento di Fisica e Astronomia ``Ettore Majorana``, Catania, Italy
\item[$^{58}$] Universit\`a di Milano, Dipartimento di Fisica, Milano, Italy
\item[$^{59}$] Universit\`a di Napoli ``Federico II'', Dipartimento di Fisica ``Ettore Pancini'', Napoli, Italy
\item[$^{60}$] Universit\`a di Palermo, Dipartimento di Fisica e Chimica ''E.\ Segr\`e'', Palermo, Italy
\item[$^{61}$] Universit\`a di Roma ``Tor Vergata'', Dipartimento di Fisica, Roma, Italy
\item[$^{62}$] Universit\`a Torino, Dipartimento di Fisica, Torino, Italy
\item[$^{63}$] Benem\'erita Universidad Aut\'onoma de Puebla, Puebla, M\'exico
\item[$^{64}$] Unidad Profesional Interdisciplinaria en Ingenier\'\i{}a y Tecnolog\'\i{}as Avanzadas del Instituto Polit\'ecnico Nacional (UPIITA-IPN), M\'exico, D.F., M\'exico
\item[$^{65}$] Universidad Aut\'onoma de Chiapas, Tuxtla Guti\'errez, Chiapas, M\'exico
\item[$^{66}$] Universidad Michoacana de San Nicol\'as de Hidalgo, Morelia, Michoac\'an, M\'exico
\item[$^{67}$] Universidad Nacional Aut\'onoma de M\'exico, M\'exico, D.F., M\'exico
\item[$^{68}$] Institute of Nuclear Physics PAN, Krakow, Poland
\item[$^{69}$] University of \L{}\'od\'z, Faculty of High-Energy Astrophysics,\L{}\'od\'z, Poland
\item[$^{70}$] Laborat\'orio de Instrumenta\c{c}\~ao e F\'\i{}sica Experimental de Part\'\i{}culas -- LIP and Instituto Superior T\'ecnico -- IST, Universidade de Lisboa -- UL, Lisboa, Portugal
\item[$^{71}$] ``Horia Hulubei'' National Institute for Physics and Nuclear Engineering, Bucharest-Magurele, Romania
\item[$^{72}$] Institute of Space Science, Bucharest-Magurele, Romania
\item[$^{73}$] Center for Astrophysics and Cosmology (CAC), University of Nova Gorica, Nova Gorica, Slovenia
\item[$^{74}$] Experimental Particle Physics Department, J.\ Stefan Institute, Ljubljana, Slovenia
\item[$^{75}$] Universidad de Granada and C.A.F.P.E., Granada, Spain
\item[$^{76}$] Instituto Galego de F\'\i{}sica de Altas Enerx\'\i{}as (IGFAE), Universidade de Santiago de Compostela, Santiago de Compostela, Spain
\item[$^{77}$] IMAPP, Radboud University Nijmegen, Nijmegen, The Netherlands
\item[$^{78}$] Nationaal Instituut voor Kernfysica en Hoge Energie Fysica (NIKHEF), Science Park, Amsterdam, The Netherlands
\item[$^{79}$] Stichting Astronomisch Onderzoek in Nederland (ASTRON), Dwingeloo, The Netherlands
\item[$^{80}$] Universiteit van Amsterdam, Faculty of Science, Amsterdam, The Netherlands
\item[$^{81}$] Case Western Reserve University, Cleveland, OH, USA
\item[$^{82}$] Colorado School of Mines, Golden, CO, USA
\item[$^{83}$] Department of Physics and Astronomy, Lehman College, City University of New York, Bronx, NY, USA
\item[$^{84}$] Michigan Technological University, Houghton, MI, USA
\item[$^{85}$] New York University, New York, NY, USA
\item[$^{86}$] University of Chicago, Enrico Fermi Institute, Chicago, IL, USA
\item[$^{87}$] University of Delaware, Department of Physics and Astronomy, Bartol Research Institute, Newark, DE, USA
\item[] -----
\item[$^{a}$] Max-Planck-Institut f\"ur Radioastronomie, Bonn, Germany
\item[$^{b}$] also at Kapteyn Institute, University of Groningen, Groningen, The Netherlands
\item[$^{c}$] School of Physics and Astronomy, University of Leeds, Leeds, United Kingdom
\item[$^{d}$] Fermi National Accelerator Laboratory, Fermilab, Batavia, IL, USA
\item[$^{e}$] Pennsylvania State University, University Park, PA, USA
\item[$^{f}$] Colorado State University, Fort Collins, CO, USA
\item[$^{g}$] Louisiana State University, Baton Rouge, LA, USA
\item[$^{h}$] now at Graduate School of Science, Osaka Metropolitan University, Osaka, Japan
\item[$^{i}$] Institut universitaire de France (IUF), France
\end{description}

\noindent Email: \href{mailto:spokespersons@auger.org}{spokespersons@auger.org}

\fbox{
\begin{minipage}[t]{0.9\linewidth}
  \textbf{Abstract:}

The Pierre Auger Collaboration has embraced the concept of open access to their research data since its foundation, with the aim of giving access to the widest possible community. A gradual process of release began as early as 2007 when 1\% of the cosmic-ray data was made public, along with 100\% of the space-weather information. In February 2021, a portal was released containing 10\% of cosmic-ray data collected by the Pierre Auger Observatory from 2004 to 2018, during the first phase of operation of the Observatory. The Open Data Portal includes detailed documentation about the detection and reconstruction procedures, analysis codes that can be easily used and modified and, additionally, visualization tools.   
  Since then, the Portal has been updated and extended. In 2023, a catalog of the highest-energy cosmic-ray events examined in depth has been included. A specific section dedicated to educational use has been developed with the expectation that these data will be explored by a wide and diverse community, including professional and citizen scientists, and used for educational and outreach initiatives. This paper describes the context, the spirit, and the technical implementation of the release of data by the largest cosmic-ray detector ever built and anticipates its future developments. 
\end{minipage}
}

\clearpage

\section{Introduction}\label{sec:intro}

During almost 20 years of data taking, the Pierre Auger Observatory~\cite{paoNIM}, the largest facility for the measurement of ultra-high energy cosmic rays (UHECR), has detected more than 20\,000 cosmic-ray events per year with an energy above 2.5~EeV (1 EeV = $10^{18}$ eV) providing, with unprecedented statistics and precision, major breakthroughs in the field.

The Observatory is located on a high-altitude plain near Malarg\"ue, Mendoza Province, Argentina, at a mean altitude of about 1400 m, corresponding to an atmospheric overburden of about 875 g\,cm$^{-2}\,$. The site lies between latitudes 35.0$^{\circ}$S and 35.3$^{\circ}$S and between longitudes 69.0$^{\circ}$\,W and 69.4$^{\circ}$\,W.  The Observatory combines two detection techniques for the detection of extensive air showers (EAS). A Surface Detector array of 1600 water-Cherenkov detectors (WCD) in a 1500 m triangular grid, SD-1500, yielding an effective area of 3000 km$^2$, provides lateral sampling of the EAS at the ground. A Fluorescence Detector (FD) comprising 24 telescopes grouped at four sites (eyes) overlooks the array and detects the UV fluorescence light emitted by the de-excitation of the nitrogen molecules previously excited by the charged particles from the EAS. 

\begin{figure*}[hbt]%
\centering
\includegraphics[width=\textwidth]{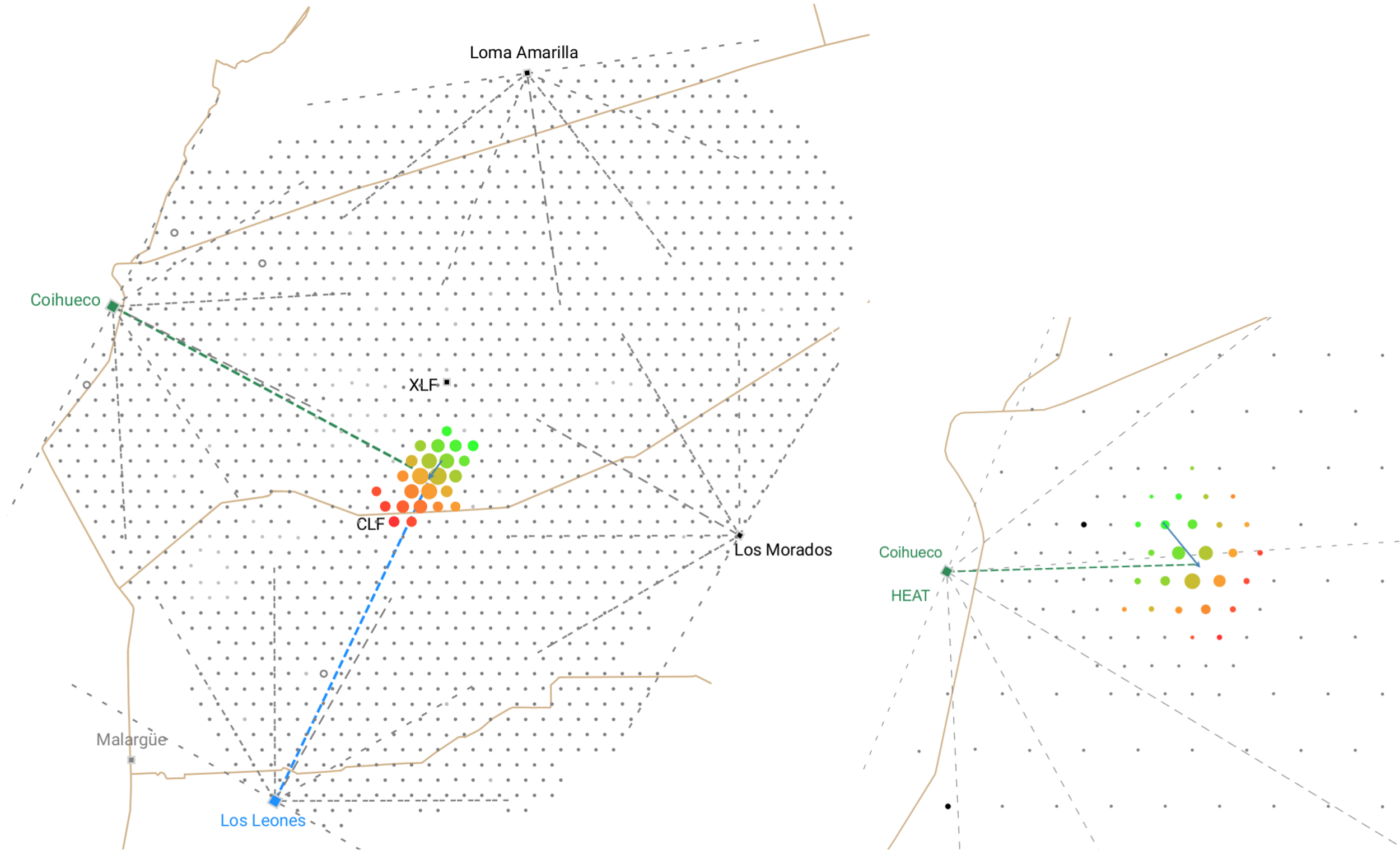}%
\caption{Map of the Observatory: left, surface detector stations (grey dots), position of the fluorescence detector eyes and field of view of the telescopes (squares, dashed lines) along with the footprint at the ground of one of the highest-energy multi-eye events in the released data sample (color scale from green to red reflects the arrival time of the shower particles at the ground). The positions of the atmospheric monitoring systems, XLF and CLF, are also shown. Right: footprint of an event detected by the SD-750 array in combination with the HEAT-Coihueco fluorescence telescopes, same color code}
\label{figMap}
\end{figure*}

Lasers for atmospheric monitoring are located towards the center of the array, at the positions marked as CLF (central laser facility) and XLF (extreme laser facility), see Fig.~\ref{figMap}, left panel. The layout includes the low energy extension array, SD-750, with a spacing of 750 m covering 24 km$^2$, and three high elevation telescopes (HEAT) installed at the FD-Coihueco site, see Fig.~\ref{figMap}, right bottom corner.

The official data-taking started on 1 January 2004 with the engineering and the pre-production arrays of 154 surface detector stations and two operating fluorescence sites~\cite{NIM04}. The installation of the SD-1500 array was completed in June 2008. The operation of the Observatory has been stable since then and is divided into two phases. During Phase I, completed on 31 December 2021, the Observatory consisted of the original layout. In Phase II, the Observatory has been upgraded with additional detectors, such as surface detector scintillators, underground muon detectors, and radio antennas, as well as upgraded electronics added to each surface detector station~\cite{upgrade}.

The rich data harvested by the Collaboration covers different and complementary fields of research. The main goal is to reveal the nature and origin of ultra-high energy cosmic rays, which relies on measurements of the energy spectrum and mass composition and is complemented by extensive searches for anisotropy in the distribution of the arrival directions at both large and intermediate angular scales. 
Ultra-high energy cosmic rays offer the unique opportunity to investigate the nature of astrophysical sources and particle interactions in a kinematic and energy region well beyond that covered by current particle accelerators. The main results include measurements of the cosmic-ray arrival directions~\cite{aniso}, energy spectrum~\cite{spectrum}, and mass composition~\cite{xmax} along with their interpretation~\cite{combi}, the possible presence of neutral primaries such as photons and neutrinos~\cite{neutrals}, and particle physics~\cite{cross}. Moreover, its potential as a multi-messenger~\cite{multim} and multi-disciplinary~\cite{multid, scalers} observatory has been proven.

The Collaboration has embraced the concept of open access to research data from its foundation. A gradual process of releasing data began as early as 2007 when the Observatory was almost completed. A public event browser with 1\% of data from the surface detector was created, and the data have been updated every year for over 10 years~\cite{edx}.
Meant for educational purposes, that portal was the first step towards making data publicly available at a time when there was no data management plan in place, and only a few other astroparticle physics experiments were releasing data before the end of their activities.

\begin{figure*}[t!]%
\centering
\includegraphics[width=0.87\textwidth]{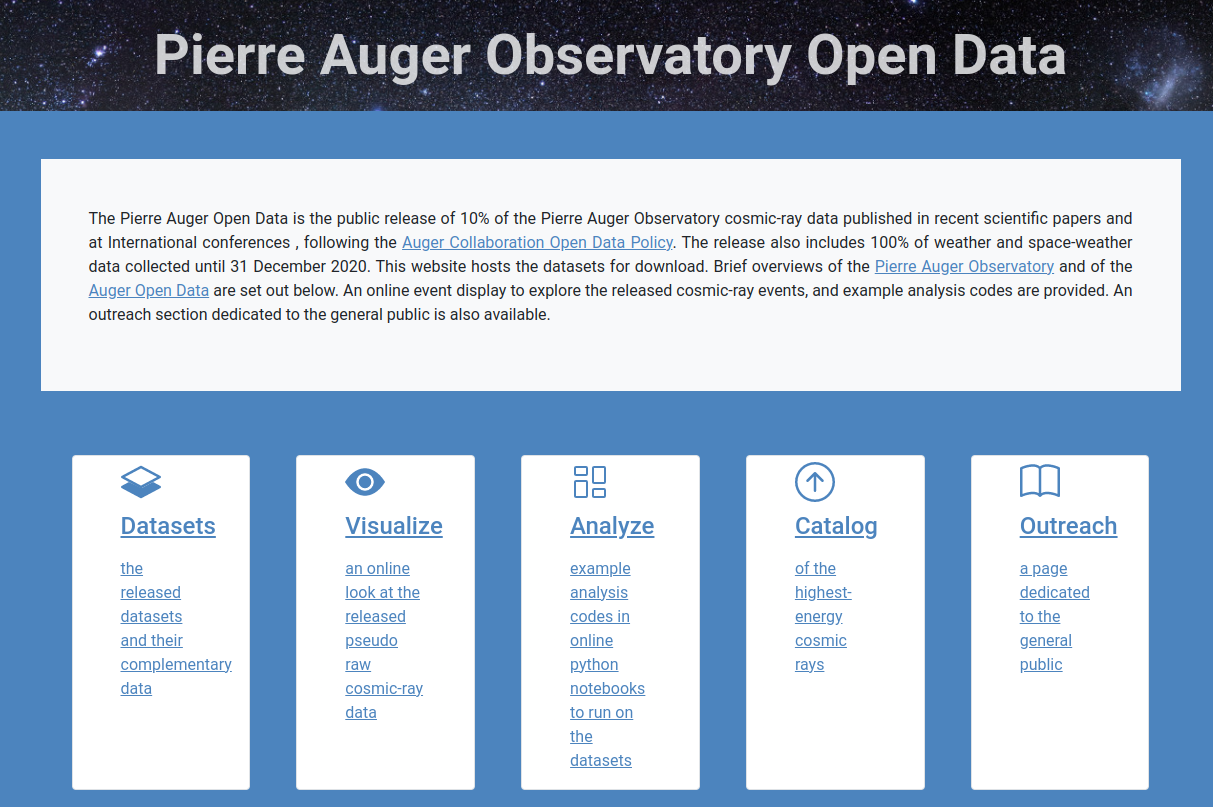}
\caption{Screenshot of the main page of the Portal~\cite{odp}, with the links to the five different sections displayed. For more details see~\cite{ICRC21, ICRC23}}
\label{fig:webp}
\end{figure*}

The Open Data Portal~\cite{odp} was set up in February 2021, towards the end of the first phase of operation of the Observatory. It contained 10\% of the cosmic-ray data used for the analyses presented in 2019 at the 36$^\mathrm{th}$ International Cosmic Ray Conference in Madison, Wisconsin, US, comprising around 25\,000 events from both the surface and the fluorescence detectors. Since then, it has been regularly updated and expanded in both the quantity and diversity of data in compliance with the \emph {Data open-access policy of the Pierre Auger Observatory}~\cite{POLICY}.

Different sets of data were added to the original sample, starting with data from atmospheric measurements, both preprocessed and raw data, and data acquired in the so called "scaler mode", recording the low-threshold counting rates of the surface detector stations. In December 2022, 10\% of cosmic-ray data above 60° was released, thus allowing to extend the aperture to a sky-coverage of 80\%. A specific outreach section was included to engage the general public in cosmic-ray physics by supporting and facilitating the use of scientific data. In March 2024, 10\% of cosmic-ray data detected by the low-energy extensions of the surface and fluorescence detectors was also released thus lowering the energy range down to 10$^{17}$ eV.

It is worthwhile to emphasize that the data from the Observatory are the result of vast and long-term human and financial investment by the international community. The Collaboration is committed to their public release and provides accompanying software tools to offer to a wider community, including professional and citizen scientists, a unique opportunity to explore and analyse the data at various levels of complexity. This is in accordance with the Berlin declaration on Open Data and Open Access, inspired by the FAIR (Findable, Accessible, Interoperable, and Reusable) principles~\cite{FAIR} for scientific data "as open as possible and as closed as necessary". The Collaboration upholds the principle that open access to the data and associated software will, in the long term, allow the full scientific potential of the data to be realised. 

This paper is conceived as follows: in Section~\ref{sec:odp} a description of the general structure and content of the Portal is given, while in Section~\ref{sec:data}, the several available datasets are described along with detailed tables with the files semantics provided in~\ref{sec:app}. In Section~\ref{sec:vis} the event visualization is introduced, and, in Section~\ref{sec:tools}, the tools for data analysis are presented. The catalog of the highest energy cosmic-ray events detected at the Observatory is described in Section~\ref{sec:catalog}. Section~\ref{sec:out} is dedicated to the outreach page and the activities connected to Open Data. Finally, in Section~\ref{sec:outro}, the experience of the Collaboration, after three years of monitoring the access and the use of the open data, is summarized, and the intentions for the future development of the open-access program are outlined. 


\section{The Portal}\label{sec:odp}

The Portal~\cite{odp, ICRC21, ICRC23} is aimed at sharing a sensible fraction of the collected data and the associated knowledge with the scientific community. This can start from either raw experimental or simulated data through reconstructed events and datasets of higher level generated by analysis workflows, all the way up to data presented in scientific publications. 

\subsection{Portal Organization} 
An introductory main page provides links to the different cards via the navigation bar. This page also displays a comprehensive overview of the Observatory, regarding the main detectors and their working principles. Appropriate references are given for further investigation.
Subsections explaining the details of the atmospheric monitoring program and the potential of the space-weather monitoring and observation of ionospheric phenomena are also provided. Finally, it includes the technical and copyright information. A screenshot of the main page is displayed in Fig.~\ref{fig:webp}.

The \emph{Datasets} card contains the list of releases, along with the available files and a detailed description of their structure and content. It also links to an exhaustive explanation of the data, and the conditions under which they were taken, reconstructed and selected, with links to the downloadable files (see Section~\ref{sec:data}).

The \emph{Visualization} card provides a user-friendly interface for selecting and browsing each of the public events by specifying an event ID or a range of reconstructed variables, such as the energy or the zenith angle. Some exemplary events, such as the highest energy or multiplicity event, are pre-selected for inspection and available through a dedicated menu (see Section~\ref{sec:vis}).  

The \emph{Analysis} card provides users with several tutorial codes to browse and analyse the data. Simple scripts for plotting histograms and graphs with the reconstructed variables and advanced routines based on the main Auger published results are available for download or can be run online (see Section~\ref{sec:tools}). 

The \emph{UHECR Catalog} card implements a browser for the 100 highest-energy events recorded by the surface detector, along with the nine highest-energy hybrid events used for their calibration~\cite{catalog} (see Section~\ref{sec:catalog}).

The \emph{Outreach} card, aimed at a wider audience, has been translated into several languages, providing a unique opportunity to share the excitement of cosmic-ray physics with the general public (see Section~\ref{sec:out}).

Finally, the Portal includes a \emph{Contact} link giving the user additional support via answers to frequently asked questions about potential technical or content-related issues that might be raised. A dedicated e-mail address for contacting the Collaboration is given.

The datasets are released under the (CC BY-SA 4.0) International License~\cite{creative}, and they have a unique Digital Object Identifier (DOI) always pointing to the current version~\cite{zenodoauger}. The user is requested to cite the general link or the specific version of the used data in any applications or publications. The DOIs of the specific released versions are detailed in Table~\ref{tab:releases}. 

\begin{table*}[hbt]
\begin{center}
\begin{tabularx}{\textwidth}{|l|l|X|X|}
\hline
\textbf{Release Tag} & \textbf{Date} & \textbf{Content} & \textbf{Specific DOI} \\ \hline
Release 3.0           & March 20, 2024    &  10\% cosmic-ray data, low energy sample &     \href{https://doi.org/10.5281/zenodo.10488964}{10.5281/zenodo.10488964}  \\ \hline
Release 2.0            & Dec 22, 2022     &  10\% cosmic-ray data, inclined sample (60$^{\circ}$\,-\,80$^{\circ}$), Outreach section with different languages &     \href{https://doi.org/10.5281/zenodo.6867688}{10.5281/zenodo.6867688}   \\ \hline
Release 1.1            & Oct 26, 2021     &  100\% atmospheric data and scaler data &     \href{https://doi.org/10.5281/zenodo.5588460}{10.5281/zenodo.5588460} \\ \hline
Release 1.0            & Feb 15, 2021     &  10\% cosmic-ray data, vertical sample (0$^{\circ}$\,-\,60$^{\circ}$) and auxiliary files, ready-to-use event display, analysis examples &     \href{https://doi.org/10.5281/zenodo.4487613}{10.5281/zenodo.4487613}   \\  \hline
\end{tabularx}
\vspace{0.3 cm}
\caption{\label{tab:releases} List of specific releases on the Open Data Portal~\cite{odp} with the corresponding DOIs}
\end{center}
\end{table*}

\section{Datasets}\label{sec:data}

Different types of data are provided via the Portal including cosmic-ray data, weather and space-weather data and other resources. Detailed explanations of the datasets, and the conditions under which data were collected and selected, are provided in the \emph{Datasets} page together with the description of the available files and of all the data fields. Tables with the files semantics are given in~\ref{sec:app}.

\subsection{Cosmic-ray data}\label{subsec31}

The cosmic-ray dataset comprises in total 81\,121 showers 3348 of which are hybrid events, i.e. recorded in combination with the fluorescence detectors. The data are calibrated and partially cleaned to reduce the level of detailed understanding of the detectors required. The events are subjected to the reconstruction procedures implemented in the official software~\cite{offline}. They are processed with the most up-to-date software at the time of the release and the software version is propagated in the files metadata. A set of selection criteria is applied to the detected cosmic-ray events in order to ensure an adequate sampling of the shower and the reliable performance of ground operation of the surface detector as well as of stable conditions of individual stations. 

A total of 25\,086 events measured with the SD-1500 have been selected. This set includes both vertical events, with a zenith angle less than $60^\circ$, and inclined events in the zenith interval $60^\circ - 80^\circ$. Their reconstructed energies are above 2.5 EeV and 4 EeV respectively, to guarantee operation in an energy regime where detection with the surface array is fully efficient. 
Hybrid events are selected by imposing a set of criteria on the status of the hardware, reconstruction of shower geometry, shower profile, and atmospheric quality. Furthermore, specific fiducial volume cuts are applied for different analyses (energy spectrum, calibration, composition) to achieve uniform acceptance and minimize the uncertainties on the corresponding observables, with events being flagged accordingly. 

The low energy sample comprises 54\,481 events detected with the SD-750 array and reconstructed with a zenith angle less than $40^\circ$ and energy above 0.1 EeV. Moreover, 197 low-energy hybrid events detected in combination with the HEAT-Coihueco telescopes and used for the energy calibration are also released. Further details of the currently released cosmic-ray samples are given in the specific sections of the \emph{Datasets} page.

Each event is downloadable as pre-processed data in JavaScript Object Notation (JSON) files, structuring, in a compact format, the meta-data on the reconstruction along with blocks of data dedicated to each of the detectors involved in the event, see Table~\ref{tab:crjson}.
The reconstructed parameters with their uncertainties and the list of the participating stations with the corresponding photomultiplier tubes (PMTs) traces are available. When an event is detected simultaneously with the fluorescence detector, the participating telescopes with the associated reconstructed energy-deposit profile and the signal recorded in the triggered pixels are provided. All the high-level reconstructed parameters from the surface and fluorescence detectors, such as energy, arrival direction, impact point at the ground, and depth of shower maximum ($X_{\text{max}}$), are also given in a summary comma-separated (CSV) file. Details of the files semantics are given in Table~\ref{tab:crcsv}.

Finally, auxiliary data files are also distributed, in particular for listing the positions of the surface detector stations and the angular field of view of the fluorescence detector pixels. The surface detector exposure and the parameters required to calculate the fluorescence detector acceptance for specific analyses are also provided. For details see Table~\ref{tab:craux}. 

\subsection{Atmospheric data}

The Observatory is a giant calorimeter, which includes the atmosphere as an important component. Therefore, a monitoring system has been set up at the detection site to measure local atmospheric parameters affecting the shower development, thus providing the knowledge required for the accurate reconstruction of observed air showers~\cite{wcorr}. Changes in the atmospheric pressure lead to changes in the rates of recorded showers. At fixed pressure, if the temperature increases, the particles in the shower will spread out more as the distance travelled between each scattering rises. Variations in  atmospheric properties also have significant effects on the rate of nitrogen fluorescence emission, as well as on the transmission of light.

The files produced by the atmospheric monitoring system include values of temperature, pressure, humidity, and wind speed recorded every five or ten minutes by five weather stations located at each fluorescence detector site and at the center of the array. In addition, the data set used to calculate the weather correction of the energy estimator derived from the surface detector, included in the standard reconstruction procedure, has also been released. The corresponding file, obtained by merging data from the weather stations, also contains the average value of the air density. Details about atmospheric data files are displayed in Table~\ref{tab:atmo}.

\subsection{Space-weather data}

Measurements of the background flux of secondary particles, produced mainly by low-energy cosmic rays (primary energies from 10 GeV to a few TeV) can be performed at the Observatory by exploiting scaler-mode data. The scaler mode is a particle counter mode, which is implemented for all the detectors of the surface array. These data are recorded for every station each second, reaching typical counting rates of 3 MHz. The temporal behavior of the number of counts is modulated by terrestrial and extraterrestrial phenomena and can thus be employed, for example, in studies of solar transient events like Forbush decreases and identification of modulations related to the solar cycle~\cite{scalers}. From September 2005, the scaler mode has been used to count, in each of the 1600 detectors, the number of times the amplitude of the signals satisfies threshold conditions corresponding to an energy deposit of between 15 MeV and 100 MeV. The current data consists of more than $10^{15}$ signal counts detected until December 2020. Details about the scalers file content are given in Table~\ref{tab:scaler}.


\section{Visualization}\label{sec:vis}

The characteristics of any shower in the dataset can be browsed via the \emph{Visualization} page. It is possible to inspect the most interesting events from a menu, such as the highest-energy events, the highest multiplicity events, and multi-eye hybrid events. In addition, a browser allows to select events by their energy, zenith angle, number of triggered stations, and arrival time. Once an event is selected, its components can be browsed using different tabs. The event files can also be directly downloaded and further processed with the provided tutorial codes.

\begin{figure*}[hbt]%
\centering
\includegraphics[width=1\textwidth]{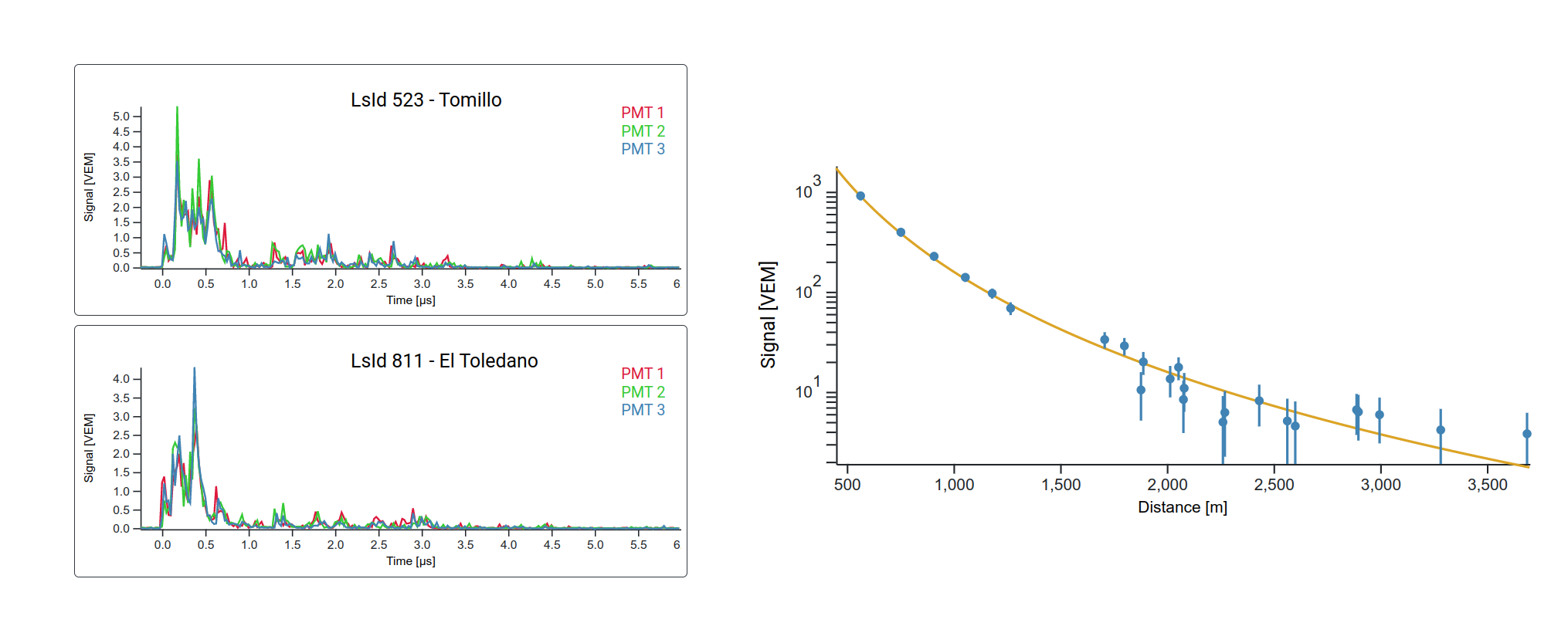}
\caption{Visualization: the exemplary event, id 081847956000 (see text for details). Left panel: FADCs traces of the PMTs signals in two WCD stations participating in the event. Right: fall-off of the signal as a function of distance from the shower axis, the so-called lateral distribution function (LDF)} \label{fig:SD}
\end{figure*}

\begin{figure*}[hbt]%
\centering
\vspace{0.3 cm}
\includegraphics[width=1\textwidth]{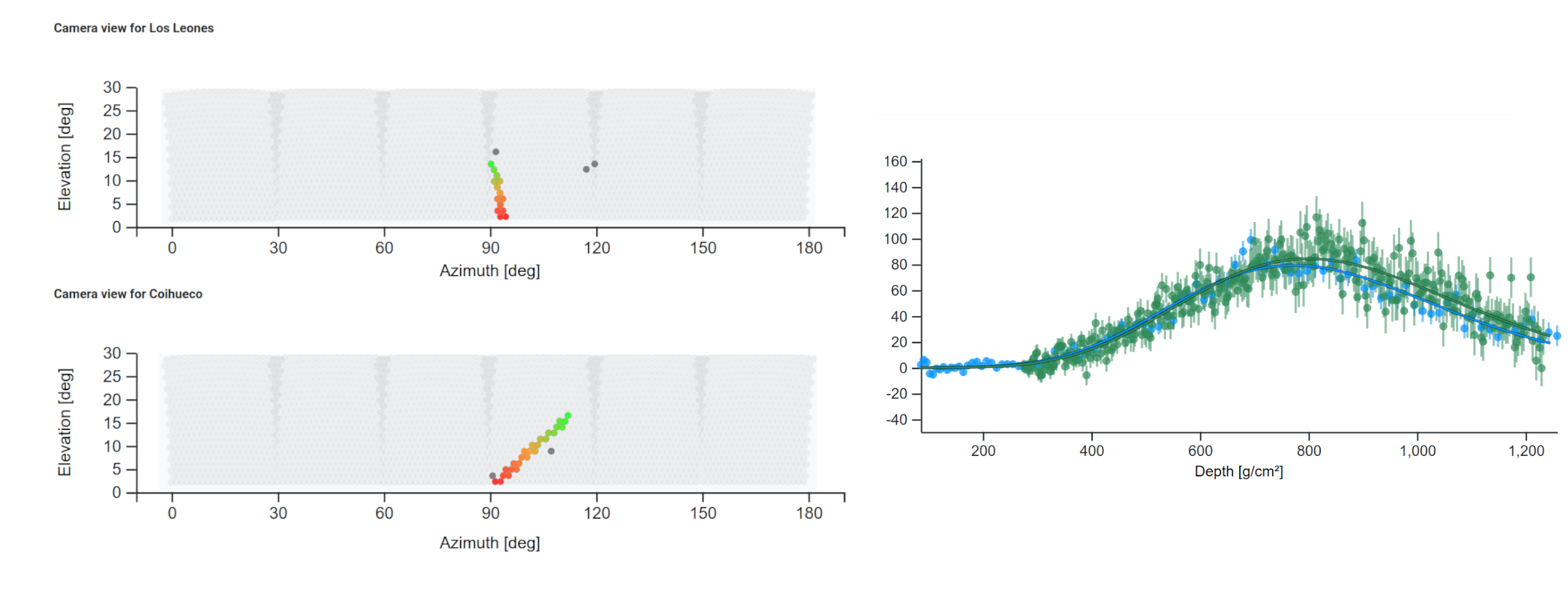}
\caption{Visualization: the exemplary event, id 081847956000 (see text for details). Left panel: camera view of the fluorescence detector; the cosmic-ray shower is seen as a trace that moves along the pixels of the camera, from early (green) to late (red) pixels. Right panel: reconstructed energy deposit as a function of atmospheric depth as measured with the two telescopes participating in the event}\label{fig:FD}
\end{figure*}

\begin{figure*}[hbt]%
\centering
\includegraphics[width=0.90\textwidth]{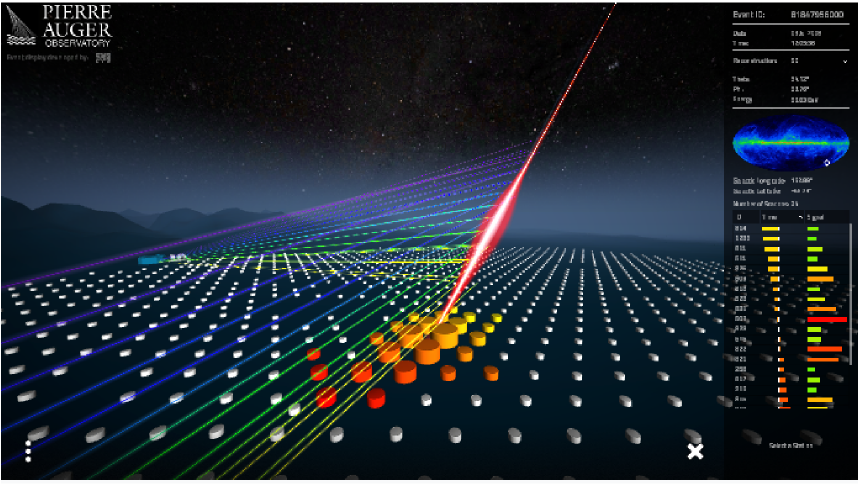}
\caption{Visualization: a screenshot of the 3-dimensional display for the exemplary multi-eye hybrid event id 081847956000 (see text for details)}\label{fig:3D}
\end{figure*}

The \textit{ground array view} tab displays a detailed map of the Observatory at the time of the event detection, showing the shower footprint at ground (color scale from green to red reflects the arrival time of the shower particles at the ground). The event shown in Fig.~\ref{figMap} with id 081847956000 is one of the highest energy multi-eye hybrid events in the released sample, with an energy of about 57 EeV and a zenith angle of 54$^\circ$. It was detected on 03 July 2008 with 24 surface detector stations and simultaneously with two sites of the fluorescence detector. Further details of this exemplary event are shown in Fig.~\ref{fig:SD}, Fig.~\ref{fig:FD}, and Fig.~\ref{fig:3D}.

The flash analog-to-digital converters (FADCs) traces of the signals of the photomultiplier tubes of all triggered stations are displayed in the \textit{SD traces} tab: the FADC traces from the PMTs of two WCD stations participating in the event are shown in Fig.~\ref{fig:SD}, left panel. 
A fit of the signal fall-off as a function of distance from the shower impact point at the ground, the so-called lateral distribution function (LDF), provides the value of the energy estimator, the signal $S(r_\text{opt})$ at an optimal distance $r_\text{opt}$ from the shower impact point on the ground in the plane perpendicular to the shower axis~\cite{SDRec}. The LDF of the exemplary event recorded by the SD-1500 array is shown in Fig.~\ref{fig:SD}, right panel. $r_\text{opt}$ equals 1000 m for the SD-1500 array and 450 m for the SD-750 array, respectively.

For hybrid events, the \emph{FD camera view} and \emph{FD reconstruction} tabs contain information from the fluorescence telescopes. The sky view of the cameras and the reconstructed energy deposit as a function of atmospheric depth, the so-called longitudinal shower profile, are shown in Fig.~\ref{fig:FD}, left and right panels, respectively.
The energy deposited per unit depth in the atmosphere, d$E$/d$X$, increases at first, along with the multiplication of particles in the cascade, and then decreases as the rate of energy loss by ionisation starts to exceed that by radiative processes. This behavior gives rise to a universal profile shape~\cite{usp}, where the position of the maximum $X_{\text{max}}$ depends mainly on the primary particle type (and its energy). The integration of the profile provides a calorimetric measurement of the total energy of the primary cosmic ray~\cite{invene}. 

The \emph {3D view} tab provides an interface to a 3-dimensional view of the events from different perspectives. An interactive and immersive view of the events and access to all reconstructed values and graphs are provided. A screenshot of the three-dimensional display for the exemplary event is shown in Fig.~\ref{fig:3D}.

\section{Analysis}\label{sec:tools}

The Open Data can be analyzed using Python Jupyter Notebooks~\cite{jupyter}. Examples are provided in the Portal, together with a tutorial introducing the Python programming language and its use with the Open Data. These Notebooks are mostly designed to require only the core Python analysis packages and can be downloaded or run online in a web browser via the Kaggle platform~\cite{kaggle}.

Each cosmic-ray event can be processed via a simple Notebook exploring the content of the associated JSON file. Depending on the file structure and the data source, different plots can be produced, such as the PMTs signals of each surface detector station, the shower footprint at the ground, or the reconstructed profile of the energy deposited in the atmosphere.

Tutorials have been developed to show how to read the CSV summary files and the JSON files and produce plots using both pseudo-raw and higher-level data. The examples demonstrate how to produce histograms and plot the trend of variables as a function of time or energy, how to produce maps of the array and of arrival directions in the sky, and how to correlate the values of two variables. More advanced analysis codes are simplified re-implementations of parts of analyses published by the Collaboration, as detailed below.

The example analyses use the most-updated version of the Auger data sets and software, which may differ slightly from those used for associated publications because of improvements to the reconstruction and calibration procedures. The codes provided maintain the spirit of the original analyses and provide insights as to how the results were obtained. They are simplified by omitting some more advanced analysis details that the user can find in the published papers. 
Even if the statistical significance is reduced with respect to what can be achieved with the full data set, the number of events is comparable to what was used in some of the first scientific publications by the Collaboration, see for example~\cite{fluxPRL}.

\begin{figure*}[hbt]%
\centering

\includegraphics[width=0.42\textwidth]{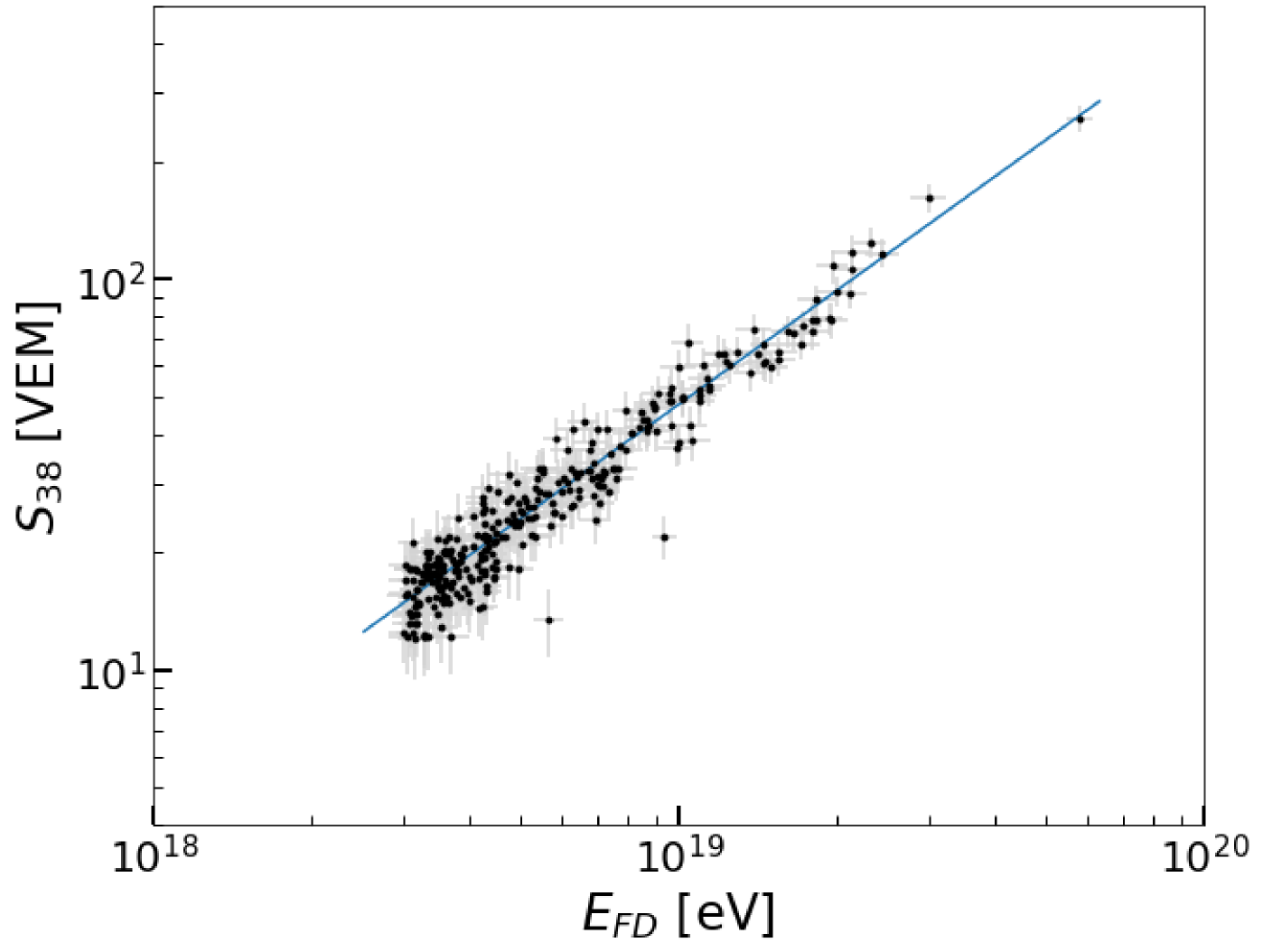}
\hspace{0.2 cm}
\includegraphics[width=0.50\textwidth]{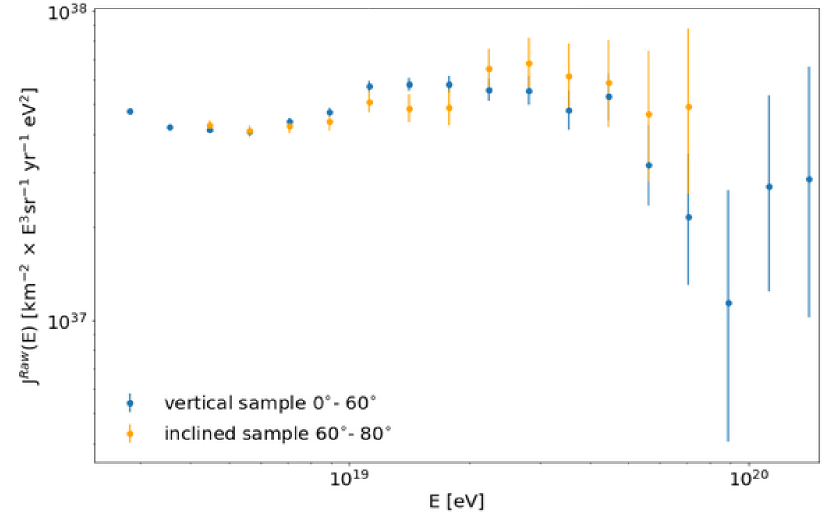}
\vspace{0.5 cm}
\includegraphics[width=0.48\textwidth]{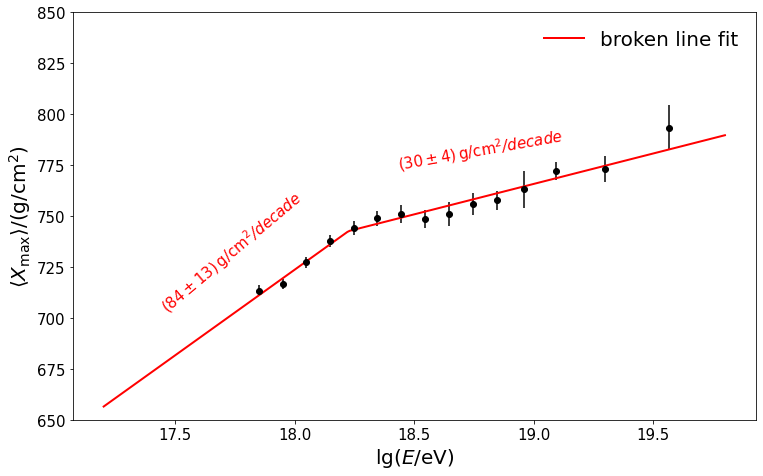}
\includegraphics[width=0.48\textwidth]{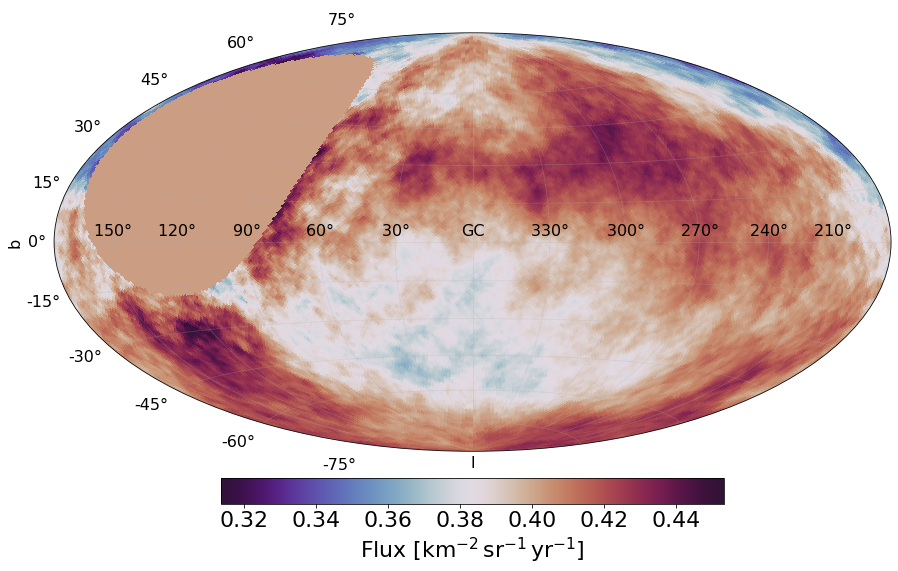}
\caption{Analysis tab: exemplary plots resulting from the Python notebooks provided. Top left panel: energy calibration of the SD energy estimator ($S_{38}$ for the SD-1500 array); top right panel: energy spectrum for vertical and inclined sample; bottom left panel: depth of shower maximum as a function of primary energy; bottom right panel: flux map for events with energy above 8 EeV}\label{figNB}
\end{figure*}

{\bf The energy calibration:} The energy estimation for the vertical events recorded using the surface detector relies on the calibration of the energy estimator, 
$S(r_\text{opt})$. $S(r_\text{opt})$ is first corrected for the the zenith-dependent attenuation, deriving the signal that a shower would have produced when coming from the median of the zenith angle distribution ($38^\circ$ for the SD-1500 and $35^\circ$ for the SD-750 sample). The calibration of the estimator is then performed by exploiting the calorimetric measurement of the energy made with the fluorescence detector for a sub-sample of high-quality hybrid events~\cite{spectrum}. The correlation between the SD estimator and the energy measured with the fluorescence detector, $E_{\text{FD}}$, is plotted in Fig.~\ref{figNB}, top left panel. 

{\bf The energy spectrum:} The estimation of the energy spectrum of cosmic rays detected with the surface detector is derived by counting the number of observed showers in differential energy bins and dividing it by the detector exposure. The bin size is constant in the logarithm of the energy, and the bin width corresponds approximately to the energy resolution. The energy threshold is fixed at  2.5 EeV for vertical events (zenith angle $< 60^\circ$), 4 EeV for inclined events ($60^\circ <$ zenith angle $< 80^\circ$) and 1 EeV for the low energy sample, as this is the energy above which the surface detector acceptance becomes independent of the mass and energy of the primary cosmic ray. The energy spectra are shown in Fig.~\ref{figNB}, top right panel. Further details are given in~\cite{spectrum, inclined, spinf}.

{\bf The depth of the shower maximum:} The estimation of the atmospheric depth at which the deposited energy for a cosmic-ray shower as a function of atmospheric depth reaches its maximum relies on the reconstruction of the longitudinal profile of events measured by the fluorescence detector, and at least one coincident surface detector station (hybrid events)~\cite{xmax}. The $X_{\text{max}}$ distributions in differential energy bins above 1 EeV for events with a zenith angle less than $75^\circ$ are built, and the energy dependence of their mean and standard deviation is derived. These can be compared to those obtained from simulations of showers produced by proton and iron primaries. The rate of change of the mean $X_{\text{max}}$ per decade of energy, the so-called elongation rate, is indicated in Fig.~\ref{figNB}, bottom left panel.

{\bf The measurement of the p-air cross-section:} The proton-air cross section for particle production at a center-of-mass energy per nucleon of \mbox{57 TeV} can be estimated by studying the shape of the distribution of $X_{\text{max}}$. The attenuation length of primary cosmic ray protons in the atmosphere is reflected in the tail of the distribution at very high values of $X_{\text{max}}$, which follows an exponential law~\cite{cross}. 

{\bf The UHECR sky:} The distribution of the reconstructed arrival directions of cosmic rays detected by the surface detector is studied to search for anisotropies at large angular scales~\cite{aniso}. The search is carried out by looking for non-uniformities in right ascension, as, for arrays that operate with close to 100\% efficiency, the total exposure as a function of this angle is almost constant. A search for the first harmonic modulation in right ascension is performed by applying the Rayleigh formalism~\cite{linsley}. In Fig.~\ref{figNB}, bottom right panel, the resulting smoothed flux map in Galactic coordinates for events with energies above 8 EeV is shown.

\begin{figure*}[t!]%
\vspace{0.2 cm}
\centering
\includegraphics[width=0.98\textwidth]{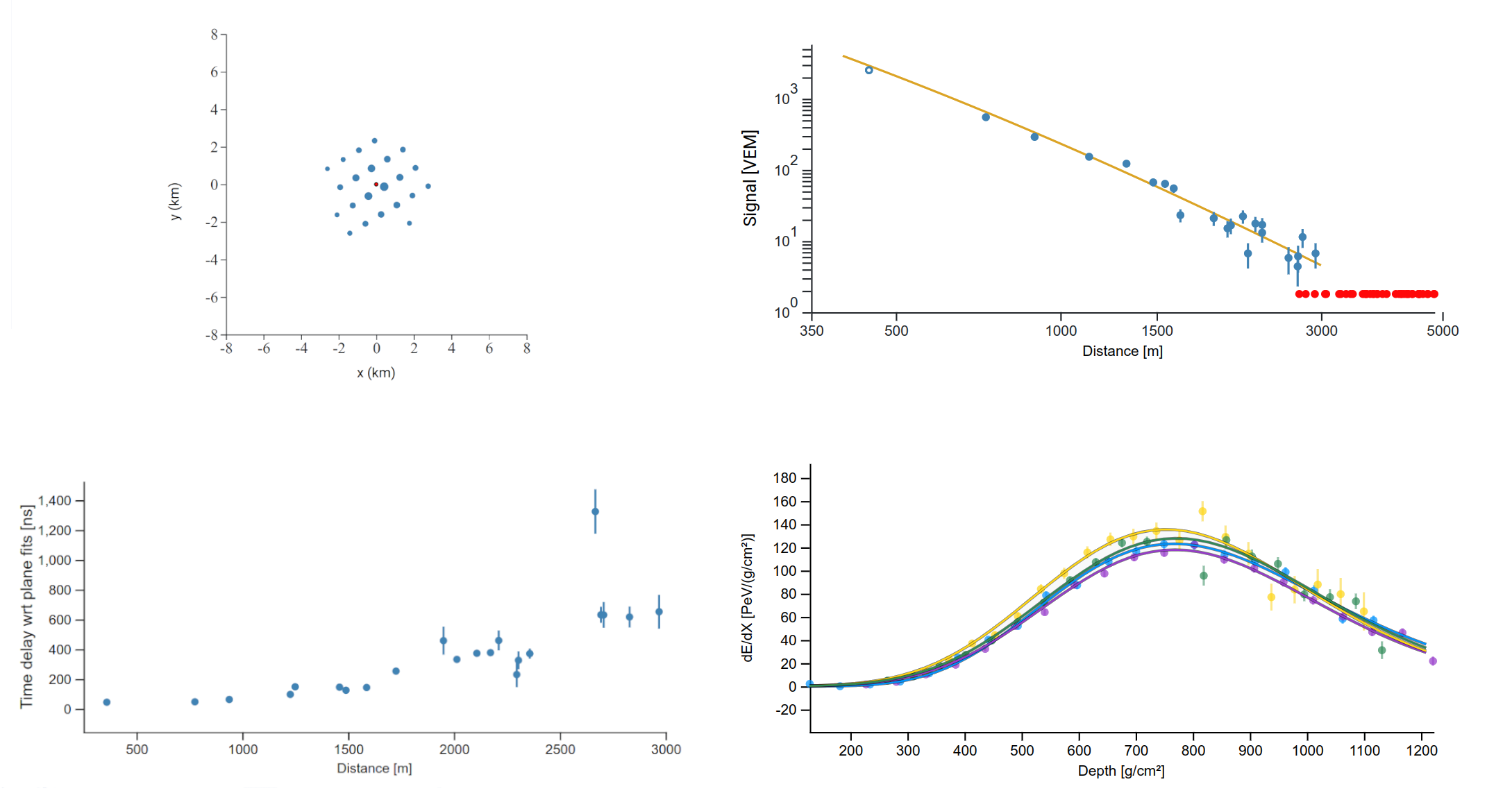}
\caption{UHECR Catalog: the highest energy multi-eye hybrid event, PAO100815 (id 102266222400). Top left panel: footprint with respect to the shower plane; top right panel: lateral distribution of the signals as a function of the distance from the shower axis. Bottom left panel: time delays of the signals with respect to a fit with a plane shower front; bottom right panel: reconstructed energy deposited in the atmosphere}\label{fig:top100}
\vspace{0.5 cm}
\end{figure*}


\section{Catalog of the highest-energy cosmic-ray events}\label{sec:catalog}

In 2023, the Pierre Auger Collaboration published a catalog of the 100 highest-energy cosmic-ray events~\cite{catalog} collected during Phase I of the data taking, along with the nine highest-energy hybrid events used for their calibration, demonstrating the quality of the data underlying the physics measurements carried out at the Observatory. In the paper, the instrumentation and the methods used to detect and reconstruct cosmic rays are also described. 

The events are available for inspection and download from the \emph{UHECR Catalog} page in the Portal. For each event, a summary of the reconstructed parameters and further details about the surface and fluorescence detector measurements are shown. The FADC traces of the photomultipliers of all triggered SD stations are also displayed in a dedicated tab.

In Fig.~\ref{fig:top100}, the main characteristics of the highest energy multi-eye hybrid event in the catalog, PAO100815 (id 102266222400), that occurred on 15 August 2010, are shown. This event had a measured energy of 82 EeV and zenith angle of 53$^{\circ}$, and it was hitting 22 surface detector stations and all four sites of the fluorescence detector. The footprint in a plane perpendicular to the shower arrival direction (top left panel) and the lateral distribution of the recorded signal as a function of the distance to the shower core in the log-log scale (top right panel) are also plotted. The user can view the time delays in nanoseconds with respect to a fit that assumes a plane shower front for the triggered stations (bottom left panel) and the reconstructed energy deposited in the atmosphere as measured by the fluorescence telescopes (bottom right panel).

\section{Outreach}\label{sec:out}

\begin{figure*}[t!]%
\centering
\includegraphics[width=0.98\textwidth]{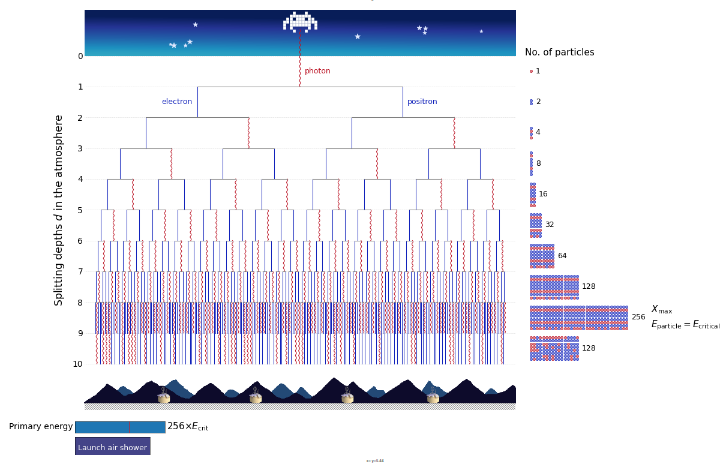}
\caption{Toy model for the development of an electromagnetic cascade in the atmosphere}\label{figEAS}
\end{figure*} 

For many years the Pierre Auger Collaboration has carried out an extensive education and outreach program for disseminating its physics results, both locally and world-wide. 

With the aim of facilitating its use by the general public, the Portal includes an \emph{Outreach} page which is a "mini-framework" built in the same spirit as the research part, but with a simplified format. The same types of data are available in the outreach framework (cosmic rays, scalers, and meteorological data) and provided in summary CSV files. The page is implementing a multilingual "in-a-nutshell" description of the physics behind UHECR research and the techniques for observing extensive air showers at the ground. An overview of the Observatory detectors and their working principles is provided, with the help of some explanatory videos, along with a simplified description of the main achievements of the Collaboration. Exemplary tutorial and analysis tools have been developed and are linked to exercises dedicated to students, teachers, and citizen scientists in the general public, providing the necessary resources to understand the data and an invitation to use them for their own inquiry by developing original education and outreach activities. Some of the exercises are described below.

{\bf Exercise using surface detector data:} The energy spectrum of cosmic-rays as detected by the surface detector is 
presented in this exercise, guiding the user to plot a histogram of the number of events detected in differential 
energy bins and normalizing it by the detector exposure. The exercise answers some of the questions the reader might have when approaching cosmic-ray physics: how the energy of a cosmic-ray primary particle relates to everyday life quantities, how rare are cosmic rays at these energies, and what is the energy carried by cosmic rays.

{\bf Exercise using hybrid data:} This Notebook provides an example of how to carry out a simple analysis of the relative composition differences between cosmic rays coming from different parts of the sky. Data collected with the surface and the fluorescence detectors simultaneously, the so-called hybrid events, can be used to extract information related to primary cosmic-ray composition. The result is shown in a sky-map in Galactic coordinates.

{\bf Exercises using non cosmic-ray data:} Two Notebooks are dedicated to non cosmic-ray data: atmospheric and space-weather data. Weather-station data are used to study the atmospheric conditions at the Observatory site. In particular, the exercise focuses on how to exploit weather stations data to calculate the value of air-density in different zones of the detection site. Data acquired via the so-called "scaler mode" can be used for space-weather science. In this Notebook the user can see how the event rate depends on the local weather conditions, such as pressure, temperature, and wind speed. 

{\bf Exercise on the shower development:} A simple toy-model for the description of the development of an extensive air shower is presented in this Notebook.  
An interactive example can be run to understand the process of particle multiplication occurring during shower development from the first interaction to reach billions of particles at the observation level. The sketch of the development of an electromagnetic cascade is shown in Fig.~\ref{figEAS}. A colour histogram of the longitudinal shower profile is plotted on the right hand-side.

\section{Outlook}\label{sec:outro}

The Pierre Auger Collaboration has been committed, since the foundation of the Observatory, to opening the access to its data. 

Data from the Observatory come from a variety of instruments and take many forms, starting from raw experimental data, through reconstructed events and datasets of higher level generated by analysis workflows all the way to data presented in scientific publications. Given the complexity and the diversity of the experimental data involved, the process of releasing them had to be started gradually. 
 
The first step towards public data dates back to 2007 with a public event browser of 1\% of cosmic-ray data. The Open Data Portal, firstly set up in February 2021 with 10\% of the cosmic-ray events, has been updated and extended by including other types of data and a catalog of the highest-energy cosmic-ray events detected with the surface and the fluorescence detectors in the first phase of operation of the Observatory.

To facilitate the continued effort of the Portal maintenance, a task for Open Data has been created under the responsibility of the Project Management and in synergy with the related physics tasks. A standardized procedure has been designed to produce the simplified and portable format for Open Data from collaboration-internal, proprietary binary files. Code libraries are maintained within the Collaboration's software repository. Open Data are processed with the most up-to-date reconstruction software. Changes are propagated into updates of the released data, and identified in the metadata of the released files by the relative software version number. The validation of the data samples and the codes to be released is performed under the supervision of analysis and detector experts and certified by the related physics tasks. Finally, the accompanying documentation follows the standard internal review for publication. 

Although only few scientific papers using the Open Data have appeared on journals or the ArXiv~\cite{arxiv} so far, the visits to the Portal, tracked via Zenodo~\cite{zenodo} and Matomo~\cite{matomo}, number more than 40\,000, while downloads of cosmic-ray samples number more than 3000 since its first publication in 2021. 
 
Open Data offer the basis for developing diverse activities dedicated to high-school and higher-level students and to the general public, focused on learning physics and enjoying programming and data analysis. 
The Collaboration actively participates in various projects to engage the general public in cosmic-ray science and the physics of the Observatory by providing open-access to cosmic-ray data and tools for astro-particle physics dissemination. In this context, we quote the International Cosmic Day and the events carried out for the International Masterclasses program, organized within the IPPOG~\cite{IPPOG} consortium, involving thousands of high school students in different countries each year. For further details, please refer to~\cite{outr23}.

In June 2023, the Collaboration Board approved the increase of the fraction of released cosmic-ray data to 30\%, planned for 2024, to mark the 20th anniversary of the official start of data taking at the Observatory, that will be applied to the whole Phase I data of the Observatory. The members of the Collaboration are convinced that this will further boost the interest in and the use of the Observatory data.  Future data from the upgraded Observatory, including new detectors, such as surface detector scintillators, underground muon detectors, and radio antennas, can be easily integrated into this framework to produce Phase II open data, to the release of which the Collaboration will undoubtedly maintain its commitment.

\clearpage
\section*{Acknowledgments}

\begin{sloppypar}
The successful installation, commissioning, and operation of the Pierre
Auger Observatory would not have been possible without the strong
commitment and effort from the technical and administrative staff in
Malarg\"ue. We are very grateful to the following agencies and
organizations for financial support:
\end{sloppypar}

\begin{sloppypar}
Argentina -- Comisi\'on Nacional de Energ\'\i{}a At\'omica; Agencia Nacional de
Promoci\'on Cient\'\i{}fica y Tecnol\'ogica (ANPCyT); Consejo Nacional de
Investigaciones Cient\'\i{}ficas y T\'ecnicas (CONICET); Gobierno de la
Provincia de Mendoza; Municipalidad de Malarg\"ue; NDM Holdings and Valle
Las Le\~nas; in gratitude for their continuing cooperation over land
access; Australia -- the Australian Research Council; Belgium -- Fonds
de la Recherche Scientifique (FNRS); Research Foundation Flanders (FWO),
Marie Curie Action of the European Union Grant No.~101107047; Brazil --
Conselho Nacional de Desenvolvimento Cient\'\i{}fico e Tecnol\'ogico (CNPq);
Financiadora de Estudos e Projetos (FINEP); Funda\c{c}\~ao de Amparo \`a
Pesquisa do Estado de Rio de Janeiro (FAPERJ); S\~ao Paulo Research
Foundation (FAPESP) Grants No.~2019/10151-2, No.~2010/07359-6 and
No.~1999/05404-3; Minist\'erio da Ci\^encia, Tecnologia, Inova\c{c}\~oes e
Comunica\c{c}\~oes (MCTIC); Czech Republic -- GACR 24-13049S, CAS LQ100102401,
MEYS LM2023032, CZ.02.1.01/0.0/0.0/16{\textunderscore}013/0001402,
CZ.02.1.01/0.0/0.0/18{\textunderscore}046/0016010 and
CZ.02.1.01/0.0/0.0/17{\textunderscore}049/0008422 and CZ.02.01.01/00/22{\textunderscore}008/0004632;
France -- Centre de Calcul IN2P3/CNRS; Centre National de la Recherche
Scientifique (CNRS); Conseil R\'egional Ile-de-France; D\'epartement
Physique Nucl\'eaire et Corpusculaire (PNC-IN2P3/CNRS); D\'epartement
Sciences de l'Univers (SDU-INSU/CNRS); Institut Lagrange de Paris (ILP)
Grant No.~LABEX ANR-10-LABX-63 within the Investissements d'Avenir
Programme Grant No.~ANR-11-IDEX-0004-02; Germany -- Bundesministerium
f\"ur Bildung und Forschung (BMBF); Deutsche Forschungsgemeinschaft (DFG);
Finanzministerium Baden-W\"urttemberg; Helmholtz Alliance for
Astroparticle Physics (HAP); Helmholtz-Gemeinschaft Deutscher
Forschungszentren (HGF); Ministerium f\"ur Kultur und Wissenschaft des
Landes Nordrhein-Westfalen; Ministerium f\"ur Wissenschaft, Forschung und
Kunst des Landes Baden-W\"urttemberg; Italy -- Istituto Nazionale di
Fisica Nucleare (INFN); Istituto Nazionale di Astrofisica (INAF);
Ministero dell'Universit\`a e della Ricerca (MUR); CETEMPS Center of
Excellence; Ministero degli Affari Esteri (MAE), ICSC Centro Nazionale
di Ricerca in High Performance Computing, Big Data and Quantum
Computing, funded by European Union NextGenerationEU, reference code
CN{\textunderscore}00000013; M\'exico -- Consejo Nacional de Ciencia y Tecnolog\'\i{}a
(CONACYT) No.~167733; Universidad Nacional Aut\'onoma de M\'exico (UNAM);
PAPIIT DGAPA-UNAM; The Netherlands -- Ministry of Education, Culture and
Science; Netherlands Organisation for Scientific Research (NWO); Dutch
national e-infrastructure with the support of SURF Cooperative; Poland
-- Ministry of Education and Science, grants No.~DIR/WK/2018/11 and
2022/WK/12; National Science Centre, grants No.~2016/22/M/ST9/00198,
2016/23/B/ST9/01635, 2020/39/B/ST9/01398, and 2022/45/B/ST9/02163;
Portugal -- Portuguese national funds and FEDER funds within Programa
Operacional Factores de Competitividade through Funda\c{c}\~ao para a Ci\^encia
e a Tecnologia (COMPETE); Romania -- Ministry of Research, Innovation
and Digitization, CNCS-UEFISCDI, contract no.~30N/2023 under Romanian
National Core Program LAPLAS VII, grant no.~PN 23 21 01 02 and project
number PN-III-P1-1.1-TE-2021-0924/TE57/2022, within PNCDI III; Slovenia
-- Slovenian Research Agency, grants P1-0031, P1-0385, I0-0033, N1-0111;
Spain -- Ministerio de Ciencia e Innovaci\'on/Agencia Estatal de
Investigaci\'on (PID2019-105544GB-I00, PID2022-140510NB-I00 and
RYC2019-027017-I), Xunta de Galicia (CIGUS Network of Research Centers,
Consolidaci\'on 2021 GRC GI-2033, ED431C-2021/22 and ED431F-2022/15),
Junta de Andaluc\'\i{}a (SOMM17/6104/UGR and P18-FR-4314), and the European
Union (Marie Sklodowska-Curie 101065027 and ERDF); USA -- Department of
Energy, Contracts No.~DE-AC02-07CH11359, No.~DE-FR02-04ER41300,
No.~DE-FG02-99ER41107 and No.~DE-SC0011689; National Science Foundation,
Grant No.~0450696; The Grainger Foundation; Marie Curie-IRSES/EPLANET;
European Particle Physics Latin American Network; and UNESCO.
\end{sloppypar}


\vspace{0.1cm}

\appendix


\section{Tables of file semantics}
\label{sec:app}

\begin{longtable}{|l|l|p{7.8cm}|}
\hline
\textbf{Section} & \textbf{Variable}  & \textbf{Range, units and description} \\
\hline
meta &  type & Name of the release \\
 &  release & Version of the release: it defines the event sample \\
 &  format & Version of data format \\
 & reconstruction software, version & Software framework used for the event reconstruction and its version \\
\hline 
info &  id & Event identification number: YYDDDSSSSSXX \\
 &  &  \quad - YY: last 2 digits of year \\
 &  &  \quad - DDD: day number between 1 and 366 \\
 &  &  \quad - SSSSS: second of the current day between 0 and 86399 \\
 &  & \quad - XX: order of the event at the current second. Time is expressed in UTC+12h, i.e., the day starting at noon \\
 &  sdid & Event number from data acquisition \\
 &  gpstime & GPS time \\
 &  date & Date and time in ISO 8601 format \\
\hline 
flags & sd1500  & [0,1] \quad 1: event is used in SD1500 array analysis \\
 &  sd750  & [0,1] \quad 1: event is used in SD750 array analysis \\
 &  hdSpectrum  & [0,1] \quad 1: event used for hybrid energy spectrum analysis \\
 &  hdCalib  & [0,1] \quad 1: event used for hybrid energy calibration analysis \\
 &  hdXmax & [0,1] \quad 1: event used for hybrid Xmax analysis \\
 &  multiEye & [0,1] \quad 1: a multi-eye event \\
\hline
fdrec &  id &  [1-6]  \quad Indicates the FD site: \\
 &  &  \quad  \quad  \quad  '1': Los Leones \\
 &  &  \quad  \quad  \quad  '2': Los Morados \\
 &  & \quad  \quad  \quad  '3': Loma Amarilla \\
 &  & \quad  \quad  \quad  '4': Coihueco \\
 &  & \quad  \quad  \quad  '5': HEAT \\
 &  & \quad  \quad  \quad  '6': HEAT-Coihueco \\
 &  gpsnanotime  & [ns]  \quad \, The GPS time of the event within its GPS second \\
 &  hdSpectrumEye  & [0,1] \quad 1: Eye used for the spectrum analysis \\
 &  hdCalibEye  & [0,1]  \quad 1: Eye used for energy calibration analysis \\
 &  hdXmaxEye  & [0,1] \quad 1: Eye used for Xmax analysis \\
 &  theta, phi&  [deg]  \quad The zenith and azimuth angles \\
 &  dtheta, dphi  &  [deg] \quad Uncertainties in zenith and azimuth angles \\
 &  l, b   & [deg] \quad Galactic longitude and latitude of the event \\
 &  ra, dec  & [deg]  \quad Right ascension and declination of the event \\
 &  totalEnergy  & [EeV] \quad Total energy of the primary particle initiating the event \\
 &  dtotalEnergy & [EeV] \quad  Uncertainty in the total energy of the event \\
&  calEnergy   & [EeV] \quad Calorimetric energy of the event \\
 &  dcalEnergy  & [EeV] \quad  Uncertainty in the calorimetric energy of the event \\
\hline
&  xmax  & [g/cm²] \quad Position of the maximum of the shower longitudinal development in the atmosphere \\
 &  dxmax  & [g/cm²] \quad Uncertainty in the position of the maximum of the shower longitudinal development in the atmosphere \\
 &  heightXmax  & [m a.s.l.]  \quad Height of Xmax above sea level\\
 &  distXmax  & [m]\quad Distance of Xmax to FD eye \\
 &  dEdXmax  & [PeV/(g/cm²)]  \quad Maximum energy deposit \\
 &  ddEdXmax  & [PeV/(g/cm²)] \quad Uncertainty in the maximum energy deposit \\
 &  x, y, z &  [m]  \quad Coordinates of the shower core projected at ground level (site coordinates system) \\
 &  dx, dy, dz &  [m]\quad Uncertainty in the coordinates of the shower core projected at ground level (site coordinates system) \\
 &  easting, northing, altitude &  [m] \quad Eastward, northward, and altitude coordinate of the shower core projected at ground level (UTM coordinates system) \\
 &  cherenkovFraction & Fraction of detected light from Cherenkov emission \\
 &  minViewAngle  & [deg] \quad Light emission angle from the shower towards the FD eye \\
 &  uspL &  [g/cm²] \quad Universal shower profile shape parameter L \\
 &  uspR & Universal shower profile shape parameter R \\
 &  duspL  & [g/cm²] \quad Uncertainty in the Universal Shower Profile parameter L \\
 &  duspR & Uncertainty in the Universal Shower Profile parameter R \\
 &  hottestStationId & Id of the SD station with the highest recorded signal \\
 &  distSdpStation   & [m] \quad  Distance of the hottest station to the shower detector plane (SDP), that includes the shower axis and the eye position\\
 &  distAxisStation  & [m] \quad Distance of hottest station to the reconstructed shower axis in the shower plane \\
\hline
eyes &  id   & [1-6] \quad Id of the FD site: \\
 &  & \quad  \quad \quad '1': Los Leones \\
 &  & \quad  \quad \quad '2': Los Morados \\
 &  & \quad  \quad \quad '3': Loma Amarilla \\
 &  & \quad  \quad \quad '4': Coihueco \\
 &  & \quad  \quad \quad '5': HEAT \\
 &  & \quad  \quad \quad '6': HEAT-Coihueco \\
 &  name & Name of the FD site \\
 &  atmDepthProf  & [g/cm²] \quad Array of slant depth points measured\\
 &  energyDepositProf & [PeV/(g/cm²)]  \quad Array of energy deposit at each slant depth, obtained from the shower profile fit\\
&  denergyDepositProf & [PeV/(g/cm²)] \quad Array of the uncertainty in the energy deposit at each slant depth, obtained from the shower profile fit \\
&  pixelID  & [1-3960] \quad Array of the pixel ids in the fluorescence site\\
\hline
&  pixelTime  & [100 ns bin]  \quad Array of the times of the signal centroid in each pixel\\
 &  pixelCharge &  [n ph]\quad Array of the light detected in each pixel (number of photons at telescope aperture)\\
 &  pixelStatus  & [0-4]\quad Array that indicates the status of the pixel \\
 &  & \quad \quad \quad 0: Background \\
 &  & \quad \quad \quad 1: Triggered \\
 &  & \quad \quad \quad 2: Pulse \\
 &  & \quad \quad \quad 3: SDP (shower detector plane) \\
 &  & \quad \quad \quad 4: TimeFit \\
 \hline
sdrec & gpsnanotime & [ns] \quad GPS time \\
 & theta, dtheta  & [deg]  \quad Zenith angle and its uncertainty\\
 & phi, dphi  & [deg]  \quad Azimuth angle and its uncertainty\\
 & energy & [EeV]  \quad Energy \\
 & denergy & [EeV]  \quad Uncertainty in the energy \\
 & l, b  & [deg]  \quad Galactic longitude and latitude \\
 & ra, dec & [deg]  \quad Right ascension and declination \\
 & x, y, z  & [m]  \quad Coordinate of the shower core (site coordinate system) \\
 & dx, dy & [m]  \quad Uncertainty in the coordinates of the shower core (site c. s.) \\
 & easting, northing, altitude & [m]  \quad Eastward, northward coordinates and altitude of the shower core (UTM coordinates system) \\
 & R  & [m]  \quad Radius of curvature of the shower \\
 & dR  & [m]  \quad Uncertainty in the radius of curvature of the shower \\
 & s1000 & [VEM]\quad Expected signal at 1000 m from the core (SD1500 array)\\
&ds1000 & [VEM]\quad Uncertainty in S(1000)\\
& s450 & [VEM]\quad Expected signal at 450 m from the core (SD750 array)\\
& ds450 & [VEM]\quad Uncertainty in S(450)\\
& s38 &[VEM]\quad Signal produced at 1000 m by a shower with a zenith angle of 38°\\
&s35 &[VEM]	Signal produced at 450 m by a shower with a zenith angle of 35°\\
&n19	&Energy estimator, N19, of a shower with a zenith angle $>$ 60°\\
&dn19	&Uncertainty in N19\\
&n68	&N19, that a shower would have produced had it arrived at 68°\\
&dn68	&Uncertainty in N68\\
&gcorr &[\%]\quad Geomagnetic correction to S(1000)\\
&wcorr &[\%]\quad Weather correction to S(1000)\\
&beta,gamma	& Slope parameters of the fitted LDF\\
&chi2	&Chi-square value of the LDF fit\\
&ndf	&Number of degrees of freedom in the LDF fit\\
&geochi2	& Chi-square value of the geometric fit\\
&geondf	& Number of degrees of freedom in the geometric fit\\
&nbstat	& Number of triggered stations used in reconstruction\\
\hline
&recstations	& List of ids of the triggered stations used in reconstruction\\
\hline
stations 		&id	&Id of the station\\
&name	&Name of the station\\
&x,y,z &[m]\quad	Coordinates of the station\\
&t &[ns]\quad Start time of the signal\\
& dt&[ns]\quad Uncertainty in the start time\\
&signalStartBin,signalStopBin	&FADC trace bins that indicate the start and stop of the signal\\
&signal &[VEM]\quad Integrated signal in the FADC traces\\
&dsignal &[VEM]\quad Uncertainty in the integrated signal\\
&sat &[0-2]\quad 0: high-gain and low-gain channels not saturated\\
& & \quad \quad \quad 1: high-gain channel saturated\\
& & \quad \quad \quad 2: high-gain and low-gain channels saturated\\
&isSelected &[0-1]\quad 1: the station is used in the reconstruction\\
&spDistance &[m]\quad Distance of the station to the core in the plane perpendicular to the shower axis (shower plane)\\
&dspDistance &[m]\quad Uncertainty in spDistance\\
&pmt1,pmt2,pmt3 &[VEM]\quad FADC traces from each photomultiplier. The length of each FADC trace is 768 bins. A bin corresponds to 25 ns\\
\hline
\caption{Cosmic-ray dataset - content of the JSON file}
\label{tab:crjson} 
\end{longtable}

\begin{longtable}{|l|l|p{6.5cm}|}
\hline
\textbf{Filename} & \textbf{Variable}  & \textbf{Range, units and description} \\
\hline
dataSummarySD1500.csv & & \\
dataSummaryInclned.csv  &id& Event identification number: YYDDDSSSSSXX \\
dataSummarySD750.csv  &  &  \quad - YY: last 2 digits of year \\
 &  &  \quad - DDD: day number between 1 and 366 \\
 &  &  \quad - SSSSS: second of the current day between 0 and 86399 \\
 &  & \quad - XX: order of the event at the current second. Time is expressed in UTC+12h, i.e., the day starting at noon \\
&sdid& Event number from data acquisition \\
&gpstime& GPS time of the eventad 1: event is used in SD1500 array analysis \\
&sd750 & [0,1] \quad 1: event is used in SD750 array analysis \\
&multiEye& [0,1] \quad 1: a multi-eye event \\
&sd\_gpsnanotime &[ns]\quad \, The GPS time of the event within its GPS second \\
&sd\_theta, sd\_phi& [deg]\quad The zenith and azimuth angles \\
&sd\_dtheta, sd\_dphi& deg]\quad Uncertainties in the zenith and azimuth angles \\
&sd\_energy& [EeV] \quad Total energy of the primary particle initiating the event \\
  &sd\_denergy& [EeV] \quad  Uncertainty in the total energy of the event \\
  \hline
 &sd\_l, sd\_b& [deg]  \quad Galactic longitude and latitude \\
&sd\_ra, sd\_dec &[deg]  \quad Right ascension and declination \\
&sd\_x, sd\_y sd\_z &  [m]  \quad Coordinates of the shower core (site coordinates system) \\
&sd\_dx, sd\_dy &  [m]  \quad Uncertainties in the x, y coordinates of the shower core \\
&sd\_easting, sd\_northing & [m]  \quad Eastward and northward coordinates of the shower core (UTM c.s.) \\
&sd\_altitude & [m]  \quad Altitude of the shower core (UTM coordinates system) \\
&sd\_R& [m]  \quad Radius of curvature of the shower \\
&sd\_dR& [m]  \quad Uncertainty in the radius of curvature of the shower \\
&sd\_s1000& [VEM]\quad Expected signal at 1000 m from the core\\
&sd\_ds1000&  [VEM]\quad Uncertainty in S(1000)\\
&sd\_s38& [VEM]\quad Signal produced at 1000 m by a shower with a zenith angle of 38°\\
&sd\_gcorr& Geomagnetic correction to S(1000)\\
&sd\_wcorr& Weather correction to S(1000)\\
&sd\_beta, sd\_gamma&  Slope parameters of the fitted LDF\\
&sd\_chi2& Chi-square value of the LDF fit\\
&sd\_ndf& Number of degrees of freedom in the LDF fit\\
&sd\_geochi2& Chi-square value of the geometric fit\\
&sd\_geondf& Number of degrees of freedom in the geometric fit\\
&sd\_nbstat& Number of triggered stations used in reconstruction\\
&fd\_id&   [1-6]  \quad Indicates the FD site \\
&fd\_gpsnanotime& [ns]  \quad \, The GPS time of the event within its GPS second \\
&fd\_hdSpectrumEye& [0,1] \quad 1: Eye used for the spectrum analysis \\
&fd\_hdCalibEye& [0,1]  \quad 1: Eye used for energy calibration analysis \\
&fd\_hdXmaxEye&  [0,1] \quad 1: Eye used for Xmax analysis \\
&fd\_theta, fd\_phi & [deg]  \quad The zenith and azimuth angles \\
&fd\_dtheta, fd\_dphi &  [deg] \quad Uncertainties in zenith and azimuth angles \\
&fd\_l, fd\_b& [deg] \quad Galactic longitude and latitude of the event \\
&fd\_ra, fd\_dec&  [deg]  \quad Right ascension and declination of the event \\
&fd\_totalEnergy& [EeV] \quad Total energy of the primary particle initiating the event \\
&fd\_dtotalEnergy& [EeV] \quad  Uncertainty in the total energy of the event \\
&fd\_calEnergy& [EeV] \quad Calorimetric energy of the event \\
&fd\_dcalEnergy& [EeV] \quad  Uncertainty in the calorimetric energy of the event \\
\hline
 &fd\_xmax&  [g/cm²] \quad Position of the maximum of the shower longitudinal development in the atmosphere \\
&fd\_dxmax&  [g/cm²] \quad Uncertainty in xmax \\
&fd\_heightXmax& [m a.s.l.]  \quad Height of Xmax above sea level \\
&fd\_distXmax& [m]\quad Distance of Xmax to FD eye\\ 
&fd\_dEdXmax& [PeV/(g/cm²)]  \quad Maximum energy deposit \\
&fd\_ddEdXmax& [PeV/(g/cm²)] \quad Uncertainty in the maximum energy deposit \\
&fd\_x, fd\_y, fd\_z& [m]  \quad Coordinates of the shower core projected at ground level (site coordinates system)\\
&fd\_dx, fd\_dy&  [m] \quad Uncertainty in the coordinates of the shower core projected at ground level (site coordinates system) \\
&fd\_easting, fd\_northing&  [m] \quad Eastward and northward coordinates of the shower core projected at ground level (UTM coordinates system) \\
&fd\_altitude&  [m] \quad Altitude of the shower core projected at ground level (UTM coordinates system) \\
&fd\_cherenkovFraction& Fraction of detected light from Cherenkov emission \\
&fd\_minViewAngle& [deg] \quad Light emission angle from the shower towards the FD\\
&fd\_uspL, fd\_duspL & [g/cm²]\quad Universal shower profile shape parameter L and its uncertainty\\ 
&fd\_uspR, fd\_duspR & Universal shower profile shape parameter R and its uncertainty\\
&fd\_hottestStationId&  Id of the SD station with the highest recorded signal (hottest)\\
&fd\_distSdpStation& [m] \quad  Distance of the hottest station to the shower detector plane\\
&fd\_distAxisStation& [m] \quad Distance of hottest station to the reconstructed shower axis\\ 
&sd\_exposure& Value of the exposure (for the SD1500 or SD750 array) at the time of the event rescaled by the data release fraction\\
\hline
\caption{Cosmic-ray dataset - content of the summary CSV files}
\label{tab:crcsv} 
\end{longtable}

\begin{longtable}{|l|l|p{7.7cm}|}
\hline
\textbf{File name} & \textbf{Variable} & \textbf{Description} \\
\hline
\emph{sdMap.csv} 
 &id & Identification number of the station \\
 & northing & [m]\quad UTC coordinates: northing \\
 & easting & [m]\quad UTC coordinates: easting \\
 & altitude & [m]\quad UTC coordinates: altitude\\
 &  start & GPS time of the first event detected by the station \\
 &  stop  & GPS time of the last event detected by the station \\
 &  sd1500 & [0,1] 1: station is part of the SD1500 array\\
 &  sd750 &  [0,1] 1: station is part of the SD750 array\\
\hline 
\emph{fdPixelMap.csv} & & \\ 
 & pixel  & [0-3959]\quad Identification number of the pixel in a FD site\\
 & eye & [1-6]\quad Identification number of the FD site\\
 & pixelTel & [1-440]\quad  Identification number of the pixel in a FD telescope\\
 &  tel & [1-6]\quad Identification number of the telescope \\
 &  col & [1-22]\quad Number of column of the pixel in the telescope \\
 &  row & [1-20]\quad Number of row of the pixel in the telescope\\
 &  backwallAngle & [deg]\quad Angle of the right wall of the FD site with respect to the East \\
 &  elevation & [deg]\quad Pointing direction of the pixel: elevation \\
 &  azimuth & [deg]\quad Pointing direction of the pixel: azimuth \\
\hline
\emph{sd1500exposure.csv},   & & \\ 
 \emph{sd1500exposureInclined.csv}, & gpstime &   GPS time \\
 \emph{sd750exposure.csv}  & sd\_exposure & [km² sr yr] \quad Value of the exposure for the SD1500 array above threshold (2.5 EeV for events below 60°, 4 EeV for events above 60°) and for the SD750 array (0.1 EeV for events  below 40°) integrated at the corresponding GPS time, rescaled for the fraction of released data\\
 & sd\_exposure\_all & [km² sr yr] \quad Full exposure for the SD1500 (SD750) array above threshold integrated at the corresponding GPS time, without rescaling for the fraction of released data\\
\hline
\emph{fdXmaxAcceptance.csv}   & & \\
 &  energyBin &  Index of the energy bin \\
 & lgMinEnergy & [log(E/eV)]\quad Start of energy bin\\
 & lgMaxEnergy & [log(E/eV)]\quad  End of energy bin\\
 &  Xacc1 & [g/cm²]\quad Xmax below which acceptance effects become relevant\\
 & Xacc1err & [g/cm²]\quad Statistical error on the parameter Xacc1\\
 & Xacc2 & [g/cm²]\quad Xmax above which acceptance effects become relevant \\
 &Xacc2err & [g/cm²]\quad Statistical error on the parameter Xacc2\\
 &lambda1 & [g/cm²]\quad Exponential slope of acceptance for Xmax $<$ Xacc1\\
 &lambda1err & [g/cm²]\quad Statistical error on lambda1 \\
 &lambda2 & [g/cm²]\quad Exponential slope of acceptance for Xmax $>$ Xacc2 \\
 & lambda2err & [g/cm²]\quad Statistical error on lambda2\\
\hline 
\emph{fdXmaxResolution.csv}   
 &  energyBin &  Index of the energy bin \\
 & lgMinEnergy & [log(E/eV)] \quad Start of energy bin\\
 & lgMaxEnergy  & [log(E/eV)] \quad End of energy bin\\
 & sigma1 & [g/cm²]\quad Width of first Gaussian\\ 
 & sigma1Err & [g/cm²]\quad Statistical error on sigma1\\
 & sigma2 & [g/cm²]\quad Width of second Gaussian \\
 & sigma2Err & [g/cm²]\quad Statistical error on sigma2\\
 & f & Relative weight between the two Gaussians\\
\hline
\caption{Cosmic-ray dataset - content of the auxiliary files}
\label{tab:craux} 
\end{longtable}

\begin{longtable}{|l|l|p{7.4cm}|}
\hline
\textbf{File name} & \textbf{Variable} & \textbf{Description} \\
\hline
\emph{weather.csv}    & & \\
&  time & [s] \quad Unix time (seconds since 1st Jan 1970, w/o leap seconds)\\
 & temperature & [°C] \quad Air temperature \\
 & pressure & [hPa]\quad Barometric pressure \\
 & density & [kg/m$^3$]\quad Air density \\
 & avgDensity2HoursBefore & [kg/m$^3$]\quad Value of air-density measured two hours earlier \\
\hline 
\emph{wsLosLeones.csv},    & & \\
\emph{wsLosMorados.csv},&  time & [s]\quad Unix time (seconds since 1st Jan 1970, w/o leap seconds) \\
 \emph{wsLomaAmarilla.csv},& temperature & [°C]\quad Air temperature \\
 \emph{wsCoihueco.csv}, & humidity &  [\%] \quad Relative humidity \\
 \emph{wsCLF.csv} & windSpeed & [km/h] \quad Average wind speed \\
 & pressure & [hPa] \quad Barometric pressure \\
 \hline
\caption{Atmospheric dataset - content of the weather stations files}
\label{tab:atmo} 
\end{longtable}

\begin{longtable}{|l|l|p{10.7cm}|}
\hline
\textbf{File name} & \textbf{Variable} & \textbf{Description} \\
\hline
\emph{scalers.csv}   & & \\
& time & [s] \quad Unix time (seconds since 1st Jan 1970, w/o leap seconds)\\
&rateCorr & [counts/m$^2$/s]\quad Corrected scaler rate\\
&arrayFraction & [\%]\quad Fraction of array in operation \\
& rateUncorr & [counts/s]\quad Average detector scaler rate, uncorrected \\
& pressure &  [hPa]\quad barometric pressure \\
\hline
\caption{Scaler dataset - content of the scaler mode file}
\label{tab:scaler} 
\end{longtable}

\end{document}